\documentclass[english,prl,twocolumn,superscriptaddress]{revtex4}
\usepackage{amssymb}
\usepackage{float}
\usepackage{color}
\usepackage{graphicx}
\usepackage{babel}
\begin{document}

\title{Cooper pair islanding model of insulating nanohoneycomb films}

\date{\today{}}
\author{S. M. Hollen}
\author{J. M. Valles, Jr.}
\affiliation{Department of Physics, Brown University, Providence, RI 02912}

\begin{abstract}

We first review evidence for the Cooper pair insulator (CPI) phase in amorphous nanohoneycomb (NHC) films.  We then extend our analysis of superconducting islands induced by film thickness variations in NHC films to examine the evolution of island sizes through the magnetic field-driven SIT. Finally, using the islanding picture, we present a plausible model for the appearance and behavior of the CPI phase in amorphous NHC films.

\end{abstract}
\maketitle

\section{Introduction} 

The appearance of a giant magnetoresistance peak near the superconductor to insulator transition (SIT) in several thin film systems\cite{Sambandamurthy:PRL2004, Kapitulnik:PhysicaC2005, Gantmakher:JETP1998, Baturina:PRL2007, Lin:PRL2011} has garnered a lot of recent attention.\cite{Dubi:PRB2006, Gantmakher:JETP1998, Galitski:PRL2005} The factors that drive its appearance are not yet understood. Of the thin film systems that exhibit this exotic behavior, nanohoneycomb (NHC) amorphous Bi thin films offer a unique morphology that might lead to an explanation.

The amorphous NHC thin films considered here were fabricated by quench condensing Sb and then Bi onto a substrate of Anodized Aluminum Oxide (AAO) (see Fig. 2a). In the zero-temperature limit, an insulator-superconductor transition can be driven by increasing the NHC film thickness, or a superconductor-insulator transition can be driven by applying a magnetic field to a superconducting film.  Both of these types of transitions bear little resemblance to their counterparts in films grown on planar glass substrates, even though we can be sure that films grown on AAO are locally homogeneous.\cite{Hollen:2011hz} Most interestingly, the insulating phase in NHC films contains localized Cooper pairs, as confirmed by the observation of Little-Parks like oscillations in the magnetoresistance.\cite{Stewart:Science2007} Recently, the localization of Cooper pairs has been attributed to the formation of superconducting islands near the SIT that develop due to nano-scale undulations in the film thickness.\cite{Hollen:2011hz} The appearance of these localized Cooper pairs is likely responsible for the puzzling features in the transport, including the giant magnetoresistance peak appearing at fields beyond the Little-Parks oscillations (1-3T).\cite{Nguyen:PRL2009} 

In this paper we examine the transport behavior of amorphous NHC films near the SIT within the context of this Cooper pair islanding model. We present images indicating how increasing film thickness or magnetic field changes the island size. Inspired by the somewhat regular arrangement of the islands in the images, we use an ordered grain array model to qualitatively explain how the evolution of island sizes can lead to the spectacular transport behavior in the Cooper pair insulator phase of these films. 

\section{Amorphous NHC film transport near the SIT}

\begin{figure*}
\begin{center}
\includegraphics[width=2\columnwidth,keepaspectratio]{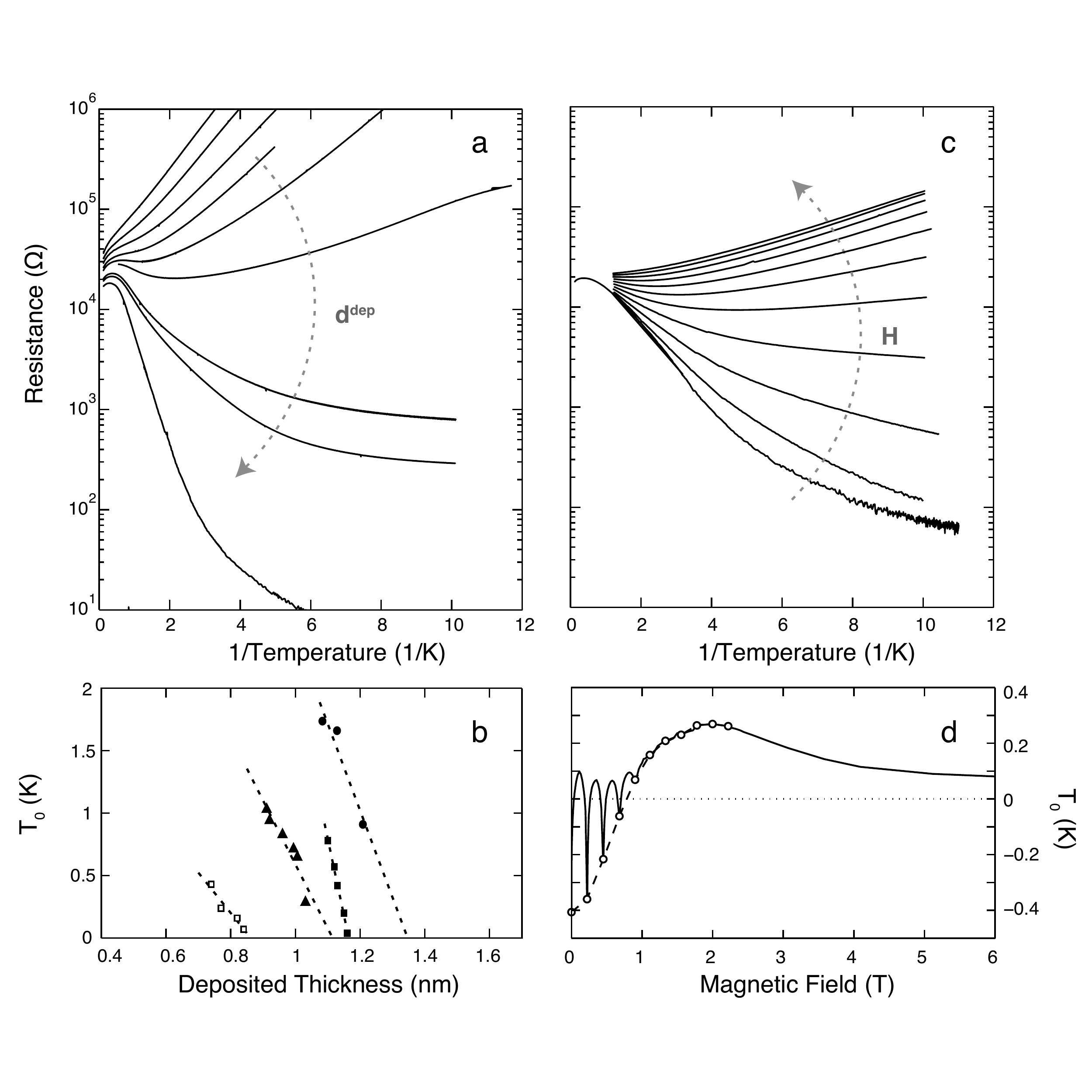}
\caption{a) Thickness-driven IST in NHC films on an Arrhenius plot.  From top to bottom, the deposited film thicknesses are 0.91, 0.92, 0.96, 0.99, 1.01, 1.03, 1.05, 1.08, 1.12 nm. b) Activation energies, obtained from the slopes of Arrhenius fits, versus deposited thickness for four different series of NHC films.  Triangles represent $T_{0}$ extracted from data in (a). c) Perpendicular field-driven SIT of a 1.13 nm NHC film on an Arrhenius plot. Field values from bottom to top are  0, 0.22,  0.45, 0.68, 0.90, 1.11, 1.33, 1.55, 1.78, 2.00, 2.22 T. d) Activation energy versus applied field for the film in (c). The $R(T)$ of the open circles (at $H=nH_{M}$, $H_{M}=0.22$ T) are shown in (c). Negative energy values are plotted for consistency; a dashed line divides negative and positive slopes (superconducting and insulating states).
\label{cap:fig1}}
\end{center}
\end{figure*}

NHC thin films on the insulating side of the SIT show simply activated behavior, where $R_{\square}(T)=R_{0}exp(T_{0}/T)$, whether they are driven by disorder or an applied magnetic field. Additionally, the activation energy, $T_{0}$, goes continuously to zero at the transition in both cases. Fig 1a shows a typical thickness-driven insulator to superconductor transition for NHC films on Arrhenius plot. The linearly decreasing activation energies of film sequences from four different experiments are shown in Fig 1b (triangles are $T_{0}$ data extracted from the series in Fig 1a). In this figure, the transition to superconductivity is marked by $d^\mathrm{dep}(T_{0}=0)$.

Complementary data for the field-driven superconductor-insulator transition are shown in Figs. 2c and d. The activation energy (Fig. 2d) changes dramatically in magnetic field, mimicking the behavior of the magnetoresistance with both oscillations and a peak in $T_{0}$.\cite{Nguyen:PRL2009,Stewart:Science2007} In contrast, the prefactor, $R_{0}$, increases monotonically by a factor of 2 through the entire field range (not shown).\cite{Nguyen:PRL2009} Because of the oscillations, the initially superconducting film undergoes 7 consecutive transitions between superconducting and insulating states (marked by crossing the $T_{0}=0$ line).  The $R(T)$ curves transversing the SIT in c are for the field values chosen at $H=nH_\mathrm{M}$ to avoid the intra-oscillation transitions (open circles in d). The form of $T_{0}(H)$ suggests two separate contributions to the activation energy: 
\begin{equation}
T_{0}(H)=T_{0}^\mathrm{osc}(H)+T_{0}^\mathrm{peak}(H).
\end{equation}
The origin of the overarching $T_{0}^\mathrm{peak}$ feature is not yet determined, but a similarly shaped feature appears in the other highly-disordered (unpatterned) film systems, mentioned in the introduction and in more detail in the discussion. 

At low fields, we expect that $T_{0}^\mathrm{osc}$ is dominated by a Josephson energy between islands, which oscillates with the matching field period, as in Josephson Junction Arrays (JJAs).\cite{Benz:PRB1988} The following expression is often used to characterize this oscillating energy scale,
\begin{equation}
T_{0}^\mathrm{osc}(H)\propto <E_\mathrm{J0}cos(\phi_{i}-\phi_{j}-2\pi A_{ij}/\Phi_{0})>,
\end{equation}
where $E_{J0}$ is the zero-field phase coupling energy, the $\phi$s are phases on neighboring islands in the film, $A_{ij}$ is line integral of the field's vector potential between islands, and $\Phi_{0}=h/(2e)$ is the magnetic flux quantum.  
The presence of this term in the activation energy is strong evidence that activated Cooper pair tunneling dominates the transport.

\section{Cooper pair islanding in NHC films}

A comparison of NHC and uniform films with increasing deposition thickness intimates that structure in the AAO substrates leads to Cooper pair localization.  The AAO is not totally flat: regular height variations appear around the holes, as can be seen in the atomic force microscope image in Fig. 2a. Since evaporated material impinges normal to the (average) plane of the substrate, thickness variations develop according to the local slope:
\begin{equation}
d(x,y)=d^\mathrm{dep}\frac{1}{\sqrt{1+(\nabla h(x,y))^{2}}},
\end{equation} 
where $d(x,y)$ is the local film thickness and $h(x,y)$ is the local height of the substrate. Evidently, the film grows thickest in the flat regions of the substrate and thinnest in the steep regions.  Fig. 2b shows the height profile along a line scan of the substrate as well as the corresponding thickness profile, calculated using Eqn. 1 and $d^\mathrm{dep}=0.89\mathrm{nm}$ (an insulating film).  Also on this plot, we mark the critical thicknesses for conduction, $d_\mathrm{Gc}^\mathrm{ref}$, and superconduction, $d_\mathrm{IST}^\mathrm{ref}$, in reference films. It is evident that some regions of the film are thick enough to support superconductivity while others are not.

\begin{figure*}
\begin{center}
\includegraphics[width=2\columnwidth,keepaspectratio]{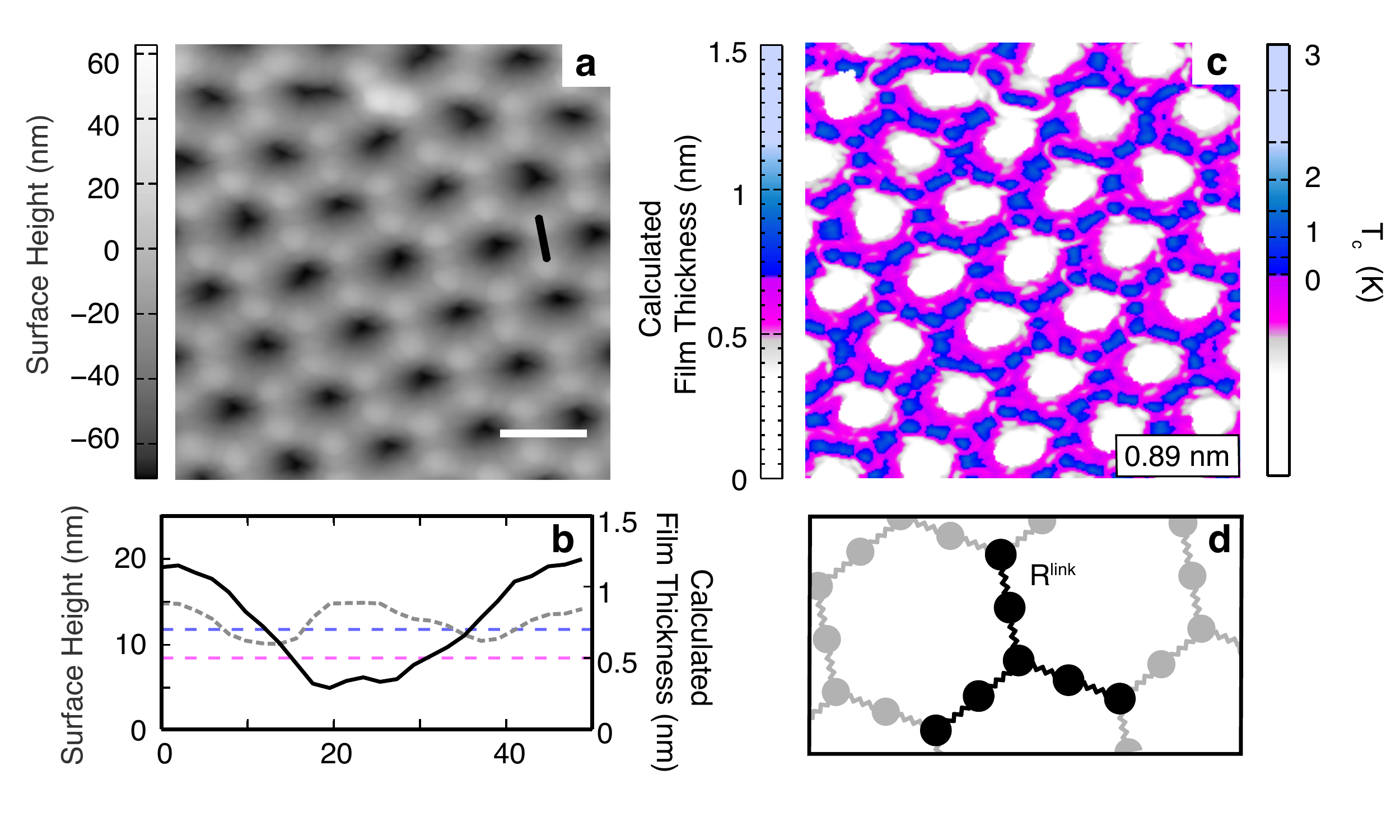}
\caption{Figure adapted from reference \cite{Hollen:2011hz} a) AFM image of the AAO substrate. The (white) scale bar spans 100 nm.  b) Height profile of substrate along a line scan (black line in (a)). The dashed gray curve is the corresponding film thickness profile calculated using Eqn. 3.  Pink and blue dashed lines mark the critical thicknesses for conductivity and superconductivity, respectively. c) Film thickness variations calculated using the image in (a) and Eqn. 3. Colored according to critical thicknesses in a reference film: pink if the film is thick enough to conduct ($d^\mathrm{dep}>d_\mathrm{Gc}^\mathrm{ref}=0.5\mathrm{nm}$), and blue if the film is thick enough to superconduct ($d^\mathrm{dep}>d_\mathrm{IST}^\mathrm{ref}=0.7\mathrm{nm}$). The right color bar defines the local coupling constant, $T_\mathrm{c}$, calculated using Eqn. 4. d) A model of the film as superconducting islands connected by resistor elements, or weak links. 
\label{cap:fig2}}
\end{center}
\end{figure*}

By applying Eqn. 1 to the AFM image of Fig 2a, we create a map of the local film thickness for a given $d^\mathrm{dep}$.  In addition, we color the map so that regions of film that conduct ($d_\mathrm{Gc}^\mathrm{ref}<d(x,y)<d_\mathrm{IST}^\mathrm{ref}$) are pink, and regions that superconduct ($d(x,y)>d_\mathrm{IST}^\mathrm{ref}$) are blue. Non-conducting film ($d(x,y)<d^\mathrm{ref}_\mathrm{Gc}$) is gray.  Fig. 2c shows the resulting morphology of an insulating film near the IST ($d^\mathrm{dep}=0.89\mathrm{nm}$). Here, regions of superconductivity are interspersed in regions of simply conducting film.

Taking the analysis one step further, we can transform the film thickness variation map into a coupling constant map, again by comparing to data of reference films. In uniform superconducting films, the transition temperature depends on film thickness as,
\begin{equation}
T_\mathrm{c}^\mathrm{ref}=T_\mathrm{c}^\mathrm{bulk}(1-d_\mathrm{IST}^\mathrm{ref}/d^\mathrm{dep}).
\end{equation}
For a-Bi, $T_\mathrm{c}^\mathrm{bulk}=6$ K and $d_\mathrm{IST}^\mathrm{ref}=0.7 \mathrm{nm}$.(Jay thesis)  Assuming this relation to be true on a microscopic level, we add a color bar to the right side of Fig. 2b to convert our map of local film thicknesses to one of local $T_{c}s$.  Considering the picture as a whole, the film appears to be made up of an array of superconducting islands connected by non-superconducting film. The local coupling constant, marked by a local $T_\mathrm{c}$, varies over an island.  It is strongest in the center, where $d(x,y)$ is largest, and weakest on the edges. By analyzing these maps through the IST, we find that the superconducting islands grow with $d^\mathrm{dep}$ and coalesce at the transition.\cite{Hollen:2011hz}    

Viewing the transitions as the coalescence of conducting or superconducting regions of film allows us to understand the critical deposition thicknesses for conduction ($d^\mathrm{NHC}_\mathrm{Gc}$) and superconduction ($d^\mathrm{NHC}_\mathrm{SC}$) in NHC films.\cite{Hollen:2011hz}  This result is corroborated by modeling the film as a network of islands linked by resistor elements, shown in Fig 2d.  At the SIT, the resistance of a single weak link in this network is consistently close to $R_{Q}=h/4e^{2}$, the critical value of sheet resistance in uniform films.  Both of these observations indicate that the weak links between superconducting islands control the transition to superconductivity in these films.\cite{Hollen:2011hz}

\section{Shrinking Cooper pair islands with an applied magnetic field}

\begin{figure*}
\begin{center}
\includegraphics[width=2\columnwidth,keepaspectratio]{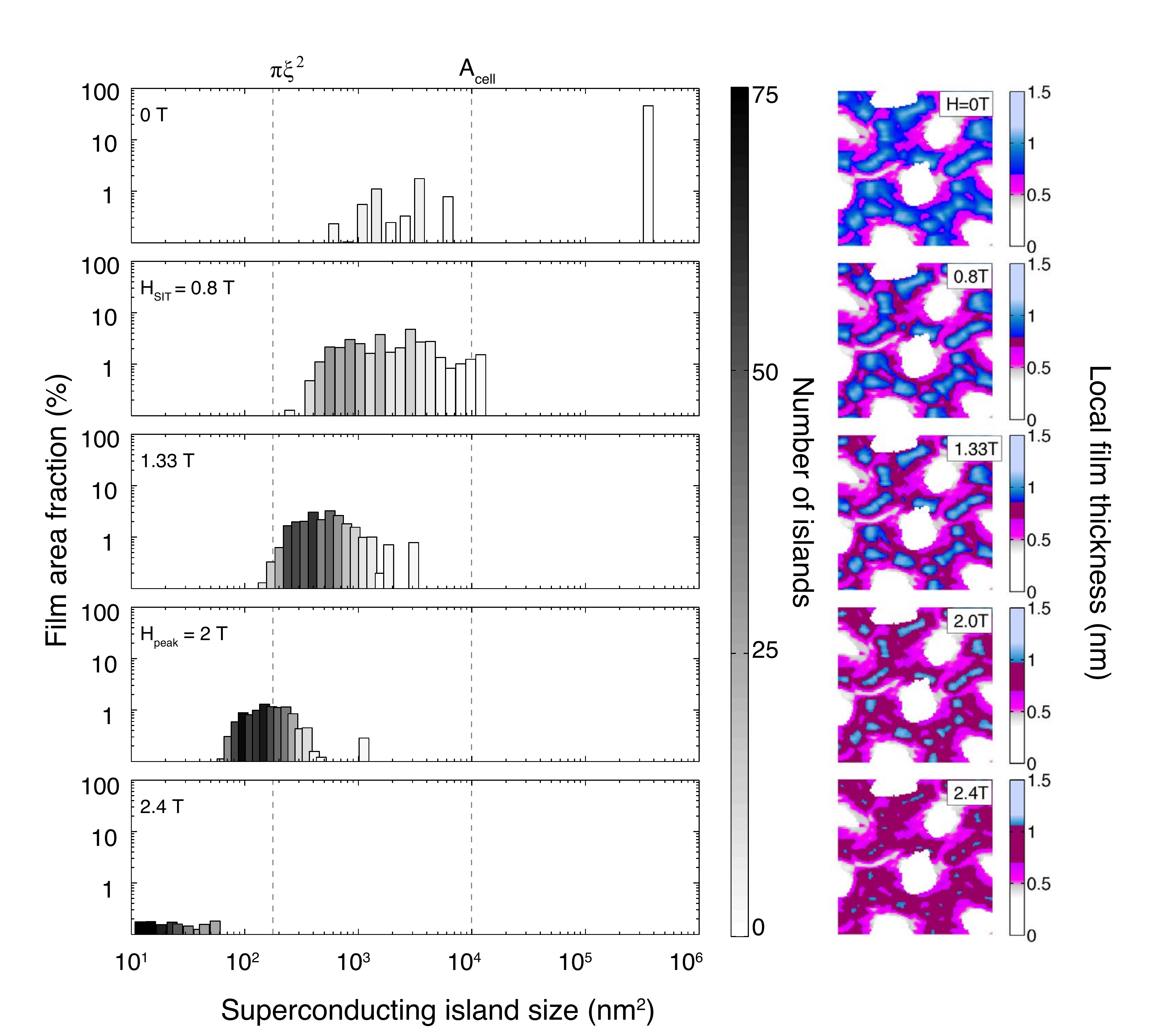}
\caption{Superconducting islands in an increasing magnetic field for the 1.08nm film of Fig 1c,d. Right: Spatial map of SC islands (blue) in an insulating background (pink) with increasing magnetic field. The cranberry color represents regions of initially superconducting film for which $H_\mathrm{c2}(d(x,y))\leq H_\mathrm{applied}$, calculated using Eqn. 5. Left: Distribution of superconducting island size and number of islands of each size for the applied field series. The first, third, and fourth panels (from the top) correspond to R(T) shown in Fig. 1c.   
\label{cap:fig3}}
\end{center}
\end{figure*}

We can extend our islanding picture to undulating films in an applied magnetic field by once again comparing the local film thicknesses to uniform films. This fairly primitive approach has led to a nice picture that qualitatively agrees with some models described in the discussion. We attribute the critical field in uniform amorphous Bi films of various thicknesses, $H_{c2}(d^\mathrm{dep})$, to local film thicknesses in undulating amorphous Bi films.  In uniform films, $H_{c2}$ depends on thickness as,
\begin{equation}
H_{c2}=7.12\mathrm{T}(1-0.7\mathrm{nm}/d).
\end{equation}
(determined by a fit to the data)\cite{Jay:thesis}.  Using this phase line, we can define a critical thickness for superconductivity for every value of applied field, $d_{SIT}(H)$. In Fig. 4, the superconducting film of Fig. 1c,d ($d^\mathrm{dep}=1.08\mathrm{nm}$) is shown for a series of applied magnetic fields.  At each field value the color bar is redefined to include a region of previously superconducting film made normal by the applied field (cranberry colored). In the figure, the islands shrink with increasing field, eventually disappearing at H=2.5T (not shown). The accompanying histograms show the magnetic field evolution of the island size distribution, as the number of islands of each size (in a 1x1$\mu$m region), and the percent film area taken up by islands of that size. At H=0T, a single island dominates the film area, but as the field is increased these islands separate.  The second panel from the top shows the rather broad distribution of islands at the field-driven SIT (0.8T), where they have just separated.  At $H_\mathrm{peak}$=2T, the islands are much smaller and fairly normally distributed in number and area around the estimated coherence area, $\pi\xi^{2}$.  Beyond the peak, at H=2.4T, the islands are all smaller than $\pi\xi^{2}$ and have nearly disappeared. 

\section{Discussion}

Dramatic transport behavior near the SIT, similar to what has been presented here (see Fig. 1), has been observed in a number of thin film systems, including indium oxide, titanium nitride, and beryllium.\cite{Sambandamurthy:PRL2004,Baturina:PRL2007,Bielejec:PRL2001}  
In particular, a number of models have been formulated to try to explain the origin of the giant magnetoresistance peak.\cite{Dubi:PRB2006, Muller:2009ea, Galitski:PRL2005, Gantmakher:JETP1998} In these models, the peak is commonly attributed to the existence of islands of incoherent Cooper pairs.  Our analysis of the morphology of NHC films suggests a well defined mechanism for island formation.  This mechanism provides a picture of the geometrical arrangement and rough sizes of the islands through the disorder and magnetic field tuned SITs. Below, we combine these unique insights with and ordered grain array model to provide a qualitative explanation for the activated insulator and MR peak.

\subsection{The hard gap at the SIT}

A number of models produce energy scales related to Cooper pairs in the insulator phase that may correspond to the activation energy seen in insulating NHC films (see Fig. 1b,d).  Trivedi and coworkers\cite{Bouadim:NatPhys2011} showed that an attractive Hubbard model with disorder predicts a pair gap that opens at the SIT and grows into the insulating state.  This gap corresponds to the energy required to inject a Cooper pair into the insulator. Very recently, M\"{u}ller\cite{Muller:2009ea,Muller:privatecomm} proposed that a mobility edge can appear in a disordered, localized bosonic system with Coulomb interactions. The distance to the mobility edge, which serves as the activation energy for Cooper pair transport, increases with disorder and magnetic field.    
Finally, and of primary interest here, an ordered array of small, Josephson coupled, superconducting grains is expected to exhibit a Mott insulator phase with an energy barrier for Cooper pair injection (see \cite{Beloborodov:RMP2007} and refs. therein).  In the limit of very weak coupling between grains this barrier corresponds to the self charging energy for a grain, $E_\mathrm{C}$. Close to the SIT, the increased coupling between grains, as measured by the Josephson energy, $E_\mathrm{J}$, leads to a downward renormalization of this barrier, $E_\mathrm{C}\rightarrow \tilde{E_\mathrm{C}}$. Deep in the insulator, $\tilde{E_\mathrm{C}}\simeq E_\mathrm{C}-E_\mathrm{J}/4$.  As the intergrain normal state resistance approaches $R_\mathrm{Q}$, this renormalization can be very substantial.(citation)  

The network of islands that results from CPs localized by coupling constant variations suggests applying the ordered grain array model to NHC films.  Within this model, we estimate that the islands contain few Cooper pairs and have a large energy level spacing.  The bare charging energy between grains, $E_\mathrm{C}$, is found to be enormous for the islands in Fig. 2 (approximately 100 meV)\cite{Stewart:Science2007} and much greater than the Josephson energy, $E_\mathrm{J}$, so that the applicability of the model would seem to rely on a very strong renormalization of $E_\mathrm{C}$ near the SIT.  

The number of Cooper pairs in a coherence volume can be estimated as $N_\mathrm{CP}=V_\mathrm{island}\nu(\epsilon_{F})\Delta$, where $V_\mathrm{island}$ is the volume of the superconducting island, $\nu(\epsilon_{F})$ is the density of states at the Fermi energy and $\Delta$ is the amplitude of the superconducting order parameter. Taking an island near the SIT to be 20nm in diameter and 1nm thick, $V_\mathrm{island}\approx100\pi\mathrm{nm}^{3}$.  We estimate $\nu(\epsilon_{F})$ for a-Bi to be $2\times 10^{22}\mathrm{cm}^{-3}\mathrm{eV}^{-1}$, the value for crystalline Pb. Finally, we estimate $\Delta$ using $2\Delta=3.5k_{B}T_{c}$ and the local maximum $T_{c}$ from the coupling constant map (Fig. 2c). Using these numbers we find that $N_{CP}\approx 2$.  Additionally, the energy level spacing in an island, $\delta=1/\nu(\epsilon_{F})V_\mathrm{island}$, yields $\delta\approx 1.76\mathrm{K}$. 

To estimate the Josephson energy, we use $E_\mathrm{J}=\pi g \Delta/2$ where the normalized conductance, $g\approx 1$ near the transition.  For $T_\mathrm{c}=2\mathrm{K}$, this expression gives $E_\mathrm{J}\approx 5\mathrm{K} (\approx 0.4\mathrm{meV})$. Finally, we calculate the charging energy with $E_\mathrm{C}=4e^{2}/\epsilon_{0}\epsilon L$ using $L=20\mathrm{nm}$, as the superconducting island size (see Fig. 2c). We find $E_\mathrm{C}\approx 10^{4} \mathrm{K}/\epsilon$, implying $E_\mathrm{C}\gg \delta, T_\mathrm{c}, E_\mathrm{J}$ for any reasonable $\epsilon$. 

Near the SIT, the condensation energy per grain, the Josephson energy, the energy level spacing, and $T_\mathrm{c}$ are all of the same order. The charging energy per grain is very large, however.  We expect it to be dominant in this model even after the necessary renormalization.  On approaching the disorder-driven SIT from the insulating side, the superconducting islands grow with $d^\mathrm{dep}$ and coalesce at the transition.\cite{Hollen:2011hz}  As the islands grow, their separation shrinks and they become better coupled.  These changes result in a diminishing $E_\mathrm{C}$ and an increasing $E_\mathrm{J}$ as $d^\mathrm{dep}\rightarrow d^\mathrm{NHC}_{IST}$. The growing islands accommodate increasingly more Cooper pairs.  These changes result in a reduction of $\tilde{E_\mathrm{C}}$, which corresponds to the decreasing $T_{0}(d^\mathrm{dep})$ in Fig. 1b.

\subsection{The magnetoresistance peak}

Here, we extend the discussion of NHC films modeled as an ordered island array to account for the effects of an applied magnetic field on their transport. With Eqn. 2, we attribute the influence of the magnetic field on the activation energy to a combination of two terms. The low field magneto-oscillations in $T_{0}^\mathrm{osc}$ (Fig. 1d) result from field-induced modulations of $E_\mathrm{J}$ and, as a result, $\tilde{E_\mathrm{C}}$.  At higher fields, $T_{0}^\mathrm{peak}$ dominates and gives rise to the magnetoresistance peak.  $T_{0}^\mathrm{peak}$ grows due to the magnetic field induced shrinking of the islands, which drives up Ec and reduces the inter-island coupling, reducing $E_\mathrm{J}$. Also, the magnetic field reduces $\Delta$, which also reduces $E_\mathrm{J}$.  Altogether, these effects drive up $\tilde{E_\mathrm{C}}$, or $T_{0}^\mathrm{peak}$.  Eventually, the magnetic field depresses $\Delta$ below $\tilde{E_\mathrm{C}}$, at which point quasiparticle tunneling rather than CP tunneling becomes energetically favorable.  As pointed out by Gantmakher \cite{Gantmakher:JETP1998} and others, the magnetoresistance becomes negative for quasiparticle tunneling because the tunneling rate increases as the field depresses the gap.

\subsection{Conclusion}

We reviewed evidence for the existence of a Cooper pair insulator phase in amorphous NHC films.  AFM images reveal that structure in the substrates induce the formation of superconducting islands separated by weak links. We presented the evolution of these island sizes with film thickness and magnetic field. We then discussed an ordered grain array model that can qualitatively account for the thickness and magnetic field tuned SITs as well as the large magnetoresistance peak observed in the Cooper pair insulator state.  

While the ordered island array scenario is plausible, experimental observations still allow other interpretations.  The grain array model attributes the activation energy to charging effects, which other models do not need to include (see \cite{Ghosal:PRL1998}).  It also requires the array to be ordered, as disorder in the array would likely lead to variable range hopping or percolation phenomena. The model of Dubi and Meir\cite{Dubi:PRB2006} invokes magnetic field induced shrinkage of superconducting islands, but also uses a percolation conduction model because many of the systems exhibiting the MR peak are disordered arrays. 

 We hope our results provide a starting point for explaining the common features in the transport across seemingly dissimilar systems exhibiting Cooper pair insulating behavior.  Future theoretical work needs to address the role played by disorder and the absence of quasiparticle transport in these Cooper pair insulators. Further experiments that can establish the importance of Coulomb interactions and disorder to these SITs are also necessary.

\section*{Acknowledgements}
This work was supported by the NSF through Grant No. DMR-0605797 and No. DMR-0907357, by the AFRL, and by the ONR. We are also grateful for support from AAUW. We thank our collaborators who contributed to the experiments highlighted in this manuscript:  M. D. Stewart, Jr., H. Q. Nguyen, J. Shainline, E. Rudisaile, and Aijun Yin.  Finally, we are grateful for discussions with N. Trivedi, I. Beloborodov, Y. Meir, M. Muller, and Z. Ovadyahu.

\section*{References}
\bibliographystyle{unsrt}
\bibliography{MRislands}

\end{document}